\documentclass[sigconf,nonacm]{acmart}

\AtBeginDocument{%
  \providecommand\BibTeX{{%
    \normalfont B\kern-0.5em{\scshape i\kern-0.25em b}\kern-0.8em\TeX}}}

\setcopyright{cc}
\copyrightyear{2024}
\acmDOI{}
\acmISBN{}

\acmConference[HCDS '24]{the 3rd Workshop on Heterogeneous Composable and Disaggregated Systems, San Diego, California, USA}{April 28, 2024}{San Diego, California, USA}

\begin{document}

\title{Leveraging Apache Arrow for Zero-copy, Zero-serialization Cluster Shared Memory}

\author{Philip Groet}
\affiliation{%
  \institution{Delft University of Technology}
  \streetaddress{Mekelweg 4}
  \city{Delft}
  % \state{Ohio}
  \country{The Netherlands}
  \postcode{2628CD}
}
\email{philip.groet@gmail.com}

\author{Joost Hoozemans}
\affiliation{%
  \institution{Voltron Data}
  \country{United States/Worldwide}
}
\email{joost@voltrondata.com}

\author{Andreas Grapentin}
\affiliation{%
  \institution{Hasso Plattner Institute, University of Potsdam, Germany}
  \country{}
}
\email{andreas.grapentin@hpi.de}

\author{Felix Eberhardt}
\affiliation{%
  \institution{IBM/Hasso Plattner Institute}
  \country{Germany}
}
\email{felix.eberhardt@ibm.com}

\author{Zaid Al-Ars}
\affiliation{%
  \institution{Delft University of Technology}
  \streetaddress{Mekelweg 4}
  \city{Delft}
  \country{The Netherlands}
  \postcode{2628CD}
}
\email{Z.Al-Ars@tudelft.nl}

\author{H. Peter Hofstee}
\affiliation{%
  \institution{IBM/TU Delft}
  \country{United States}
}
\email{hofstee@us.ibm.com}

\begin{abstract}
This paper describes a distributed implementation of Apache Arrow that can leverage cluster-shared load-store addressable memory that is hardware-coherent only within each node. 
The implementation is built on the \emph{ThymesisFlow} prototype that leverages the OpenCAPI interface to create a shared address space across a cluster. 
While Apache Arrow structures are immutable, simplifying their use in a cluster shared memory, this paper creates distributed Apache Arrow tables and makes them accessible in each node.
\end{abstract}

\keywords{Apache Arrow, ThymesisFlow, Cluster Shared Memory, Memory disaggregation, OpenCAPI, PowerPC, Parallel computing, Power9}

\maketitle

\section{Introduction}

Unlike compute resources, that can be flexibly traded off against performance, main memory in each node in a cluster must be provisioned for a worst case to achieve prevent severe performance degradation.
In addition, to prevent jobs from being inadvertently killed because they ran out of memory users often over-estimate their memory requirements.
The result is internal fragmentation and significant memory under-utilization.
A study by Google into its Borg clusters reveals that only around 40\% of memory is used~\cite{borg}, and a study by Microsoft shows 50\% of VMs never touch 50\% of their memory~\cite{pond}. 
Considering that memory is one of the largest and growing contributors to the total cost of a server~\cite{al2023memory},
it is clear that 
systems capable of sharing memory nodes across a cluster could be a major cost saver.

The ThymesisFlow~\cite{ThymesisFlow} prototype 
enables byte-addressable shared memory regions between nodes in a cluster, completely transparent to the application. 
In other software RDMA implementations such as FastSwap~\cite{fastswap}, memory is copied/swapped to local memory, while ThymesisFlow memory stays in one place and transactions are sent to that memory.
All data can be load-store accessed 
and cached locally, speeding up repeated accesses to remote memory locations.
While built leveraging the OpenCAPI interfaces, ThymesisFlow does break with the OpenCAPI goal of being fully cache coherent as 
cached instances of accessed remote memory locations will not be updated when the remote node invalidates the cache.

We propose to use the in-memory format of Apache Arrow~\cite{apachearrow}. 
Arrow allows for various applications to access the same data, without copying and without serializing any data~\cite{arrow_genomics, arrow_parquet}.
The key aspect we utilize is that most Arrow objects, once instantiated, are immutable and thus the missing cache coherency is not a problem with read-only access.
This work extends Arrow's ability to be readable by multiple \emph{applications} on the same machine, to multiple \emph{machines} connected with ThymesisFlow. 
Accessing the shared memory through the Arrow API allows ensuring memory consistency, while allowing applications to leverage the shared memory without worrying about cache coherence. Our implementation extending Apache Arrow is open-source and available on github~\cite{github_repo}.

\section{Background}\label{sec:background}

Scaling workloads can fundamentally be done in a \emph{vertical} or \emph{horizontal} fashion.
The former is done by adding resources in a single large SMP system with multiple CPU-sockets coherently interconnected.
However, this type of system has its limitations: maintaining coherence of the caches leads to performance problems due to latency and bandwidth overheads of the coherency protocol.
These limitations are addressed with \emph{horizontal scaling} where resources are available in the form of networked clusters of servers with e.g. Ethernet as the interconnect.
However, in those clusters the cost of communication is significantly higher than in SMP systems~\cite{lim2012system}. 
Consequently, workloads with tightly coupled communication are better suited for \emph{vertical scaling}, whereas the opposite is true for \emph{horizontal scaling}~\cite{bigdata_scale}.
By combining ThymesisFlow with Apache Arrow, this paper minimizes data copy bottlenecks which currently hinder efficient communication between server nodes.

\subsection{Zero-copy, zero-serialization}

In a coordinated compute cluster, the classical approach to data transfer involves a serialization step, broadcasting the data to every node through the network, and then de-serializing it into a format known to the local machine. 
These might be expensive operations, e.g. if pointers need to be resolved and converted to relative offsets.
To enhance traditional data transfer methods, a logical progression is to establish a "common language" for systems to communicate in. The goal is to have a uniform in-memory data format. 
Apache Arrow aims to achieve this standardization.

Previous work on this topic \cite{incpetion_plasma} succeeded in integrating Apache Arrow with ThymesisFlow using the Plasma object store, thereby enabling transparent data communication across multiple compute nodes. However, the use of Plasma requires extra copy operations, which reduces efficiency.

\begin{figure}
    \centering
    \includegraphics[width=\linewidth]{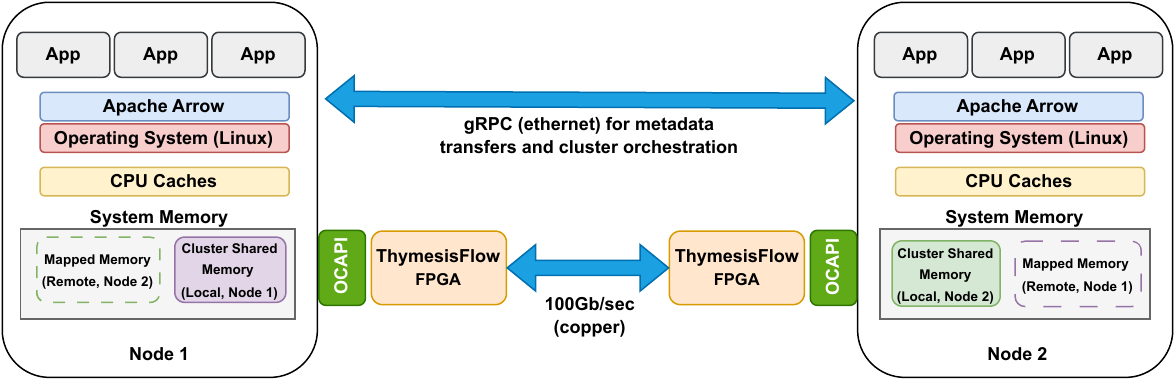}
    \caption{High-level system components. Arrow as a user library, ThymesisFlow connecting the two OpenCAPI busses together allowing for remote memory accesses.}
    \label{fig:system-diagram}
\end{figure}

This work eliminates the need for serialization and reduces the cost of copy operations. 
As shown in Figure \ref{fig:system-diagram}, this is done by leveraging the load-store based sharing of node memory across systems offered by ThymesisFlow.
A concept we call \emph{cluster shared memory} can be utilized to eliminate copy operations, introducing a \emph{zero-copy} paradigm underpinning the \emph{zero-serialization} paradigm formulated by Arrow.

\subsection{Cluster shared memory}

This paper introduces the new term \emph{cluster shared memory} (CSM) to describe a system that shares memory addresses across a cluster.
Depending on how cache coherence is enforced, we can define three levels of CSM:
\begin{itemize}
    \item Non-coherent CSM (NC-CSM): this refers to memory sharing without any coherence guarantees.
    \item Locally-coherent CSM (LC-CSM): this refers to enforcing cache coherence on individual nodes only. \emph{Our work targets LC-CSM clusters.}
    \item Globally-coherent CSM (GC-CSM): this refers to enforcing cache coherence at the cluster level.
\end{itemize}

\subsection{ThymesisFlow}

ThymesisFlow is an open-source HW/SW co-designed memory disaggregation prototype. 
A ThymesisFlow system supports memory \emph{lending} and \emph{borrowing} across nodes.
A \emph{lender} may transparently map parts of a \emph{borrower's} main memory into its local physical address space, as if additional memory modules were plugged into the \emph{lender}. It does this using an FPGA-based NICs attached to the Power9 OpenCAPI bus.
In the prototype,  
the connection between the two machines is a 100Gib/s
link, 
with an effective bandwidth up to
$\sim$10GiB/s~\cite{ThymesisFlow}
and a RTT latency of 
$\sim$650ns~\cite{openpower}.

Using the OpenCAPI connection allows for communication between connected devices in a partially cache coherent fashion.
This paper focuses on a scenario where both the \emph{lender} and \emph{borrower} access the shared memory with concurrent read and write accesses. 
When
a borrower reads from a lender's memory, first the borrower CPU cache is accessed, then the read is sent through the ThymesisFlow link, then the lender cache is snooped through the OpenCAPI bus, and finally if both caches miss, the actual memory is read. Because this setup shares architecture with local memory accesses, CPU features such as memory pre-fetching, out-of-order processing and pipelining are still active for remote memory~\cite{Yang2022}.
However, while memory is coherent within the node, it is not coherent across nodes.
In the methodology section of this paper, we outline solutions to these cache coherency issues.

In Power10 the concept of \emph{Memory Inception} was announced, which also enables disaggregated memory sharing in a cluster. 
With Memory Inception, the memory latency is expected to be significantly lower than ThymesisFlow, with 
only 50-to-100 nanoseconds of additional latency incurred through the link~\cite{Russell_2020}.
Additionally, with the OpenCAPI standard integrated into CXL~\cite{sharma2019compute, opencapi_cxl}
and CXL 3 offering memory pooling functionality among several systems,
our approach promises to inspire comparable configurations on other platforms.

\subsection{Apache Arrow}

Apache Arrow is an in-memory data format specification for tabular data organized in columnar structures, with corresponding libraries for various popular programming languages. 
It supports many different data types, including complex ones such as dictionaries, nested and variable-length data.
The power of the Arrow format lies in its in-memory data layout, which allows for interoperability of objects between different applications. It enables a Python program to create a dataset, hand the metadata to another application implemented in a different programming language, which is then able to read the data without marshaling or serialization, or even copying the data itself. 
Finally, Apache Arrow tables use offsets rather than pointers.

Apache Arrow objects, once instantiated, are immutable.
Modifying data requires creating a new object. 
The structure in which Arrow encodes arrays is called a \emph{RecordBatch}.
The structure of a RecordBatch is encapsulated in the library supporting each language, and not part of the location agnostic data format. For example, in the Arrow C++ library, the RecordBatch is implemented as a class instance, and Arrow Arrays are stored with the C++ \texttt{std::vector} type. Every node reading the data will need to instantiate its platforms Arrow library with this data.
In Section~\ref{ssec:serializing}, we describe how we serialize the structure and references of columnar data.

\section{Methodology}\label{sec:methodology}

\subsection{Serializing Arrow columnar structure} \label{ssec:serializing}

Apache Arrow already contains methods for serializing a full table, including the data itself, into a buffer, this
is done using the RecordBatchWriter and RecordBatchReader IPC methods, useful for copying objects between nodes.
However, these IPC API calls are not zero-copy and will write data into a new buffer, and move it to a shareable location. 
To be able to serialize only the table descriptor of an Arrow object, we modify the IPC API of Arrow to not include the data, only the table descriptor and a reference to the data.

The data itself does not need to be copied, as it is 
placed in a 
globally accessible shared
memory
When
another node wants to access the table, it only needs to serialize, send, and de-serialize the structure
and reference information. Because ThymesisFlow allows for creating a shared address space across nodes where every address uniquely designates memory, all the pointers to the data remain valid.
This way, no extensive redesign of Arrow is necessary, as the general structure of the code base
to instantiate objects is kept intact.

\subsection{Flushing CPU caches before invalidity}

We need to take care when creating new Arrow objects that are accessible by other nodes. When CPU caches are populated with data of a certain memory region, and another CPU writes to that region, only the writer's CPU cache will be updated. This poses a problem when the CPUs with incorrect cache entries try to read data, their request will hit only their local cache, not reading the newly updated remote memory. Thus, we need to first invalidate the caches of all CPUs, so that the caches become up to date when they query the newly updated memory. Power9 processors, however, do not support cache invalidation and data cache flush instructions must be used instead. We use the \texttt{dcbi} instruction to flush single 128-byte cachelines.

Flushing, however, may change the backing memory. 
For the purpose of emptying cache-lines we use flushing operations on all CPUs in a cluster to empty the relevant caches of memory regions. We execute the flush operations before we write the actual wanted data, allowing for the new data to then be repopulated in all the CPU caches on reads.

Power9 has an out-of-order instruction execution unit. To ensure no calculations are done before all memory blocks have been flushed, we add memory barriers before and after all flushing operations. We do not need to place the barriers between every flush operation, as we do not care if individual flush instructions are swapped.

After the initialization of an object has finished, we do no longer need to worry about cache coherency. The Apache Arrow format guarantees that created objects are immutable. 
Thus, when the data is written to the memory, and all CPU caches are made to be coherent once, any and all reads afterwards will update the CPU caches with up-to-date data. Arrow will guarantee no writes happen to the region, and we thus do not need to make the system coherent again.
The expensive invalidate operations only happen during initialization of the data, during reading of the data we have no overhead except refetching data into local CPU caches.

A typical flow for creating an Arrow object on shared memory looks like this:
\begin{itemize}
    \item Allocation: memory owning node allocates buffer and passes address to requesting node
    \item Clear cache-lines: all processors will flush their caches of the requested memory region
    \item Write into memory: any processor in the cluster writes the data to the shared memory region.
    \item Flush when not local: if the processor has written to memory it does not own, it must flush its CPU cache to ensure the data is actually written to the remote memory.
    \item Coherent reading: any processors which read this newly created memory region will not have local caches of this region. Subsequent reads will thus populate the CPU cache with up-to-date values.
\end{itemize}

\subsection{Preventing Address Translation by mapping regions to the same address on every node}

Even though the Arrow data itself does not contain any absolute pointers, the metadata does.
This includes metadata containing absolute pointers to memory where the data buffers are stored. 
Therefore to transfer information where an Arrow object is stored, all pointers to data buffers would have to be updated with the new location where data is mapped. To ease this process, we map every memory region to the exact same location in every node. 
We do this using the Linux mmap flag \texttt{MAP\_FIXED}. This flag tells the kernel that the address passed is not a suggestion, but an exact requirement where the region is mapped to. There are caveats to this, since we need to make sure on each participating process that the requested region in virtual memory is actually free and available to map, otherwise the mmap call will fail.

\subsection{Allocating in custom memory regions}

In a standard application, malloc and its family of functions only give the application limited control where memory is mapped. When a user calls malloc, the c library decides if the currently available heap space  is sufficient to satisfy the new request. If the current heap is not big enough, more pages may be requested from  the kernel using the \texttt{brk} and \texttt{sbrk} syscalls, or the c library may decide to map pages directly using mmap for large allocations. We cannot however instruct malloc to allocate in a certain memory region. 

For this reason, we extended Apache Arrow to include a memory manager. The manager allows for defining custom memory regions that are made available to malloc to satisfy allocation requests from.

\subsection{Remote memory allocations}

We choose an architecture where the CPU that owns the backing memory is responsible for handling allocations and cache behaviors. We chose this to prevent race conditions, and to have only one CPU be responsible for managing the malloc data structures. Concretely, this means that when a remote node wants to write to local memory, it will have to first request memory from the owning node. The owning node will then call local malloc methods and return the allocated address to the remote node.

This architecture allows for any node in a cluster to write to any other node. The result is an architecture where a single node writes data to all other nodes, not only nodes writing to their own local memory. Some applications are:
\begin{itemize}
    \item One node writes a dataset to all other nodes.
    \item In a big data pipeline, we can have the result of one stage be immediately written to the memory of the next stage.
\end{itemize}

\subsection{Spanning Tables across nodes}

Arrow allows for creating not only contiguous columnar data such as a RecordBatch, but also columnar data which has non-contiguous columns called Tables. Unlike RecordBatches which contain Arrays which guarantee contiguous memory buffers, Tables contain ChunkedArrays. ChunkedArrays may contain multiple contiguous Arrays, making the non-contiguous Chunked Array. Chunked Arrays are not part of the Arrow memory format, but rather is a library abstraction on top of the Array memory format spec.

With Arrow data being accessible to every other node in a cluster we can use the ChunkedArray abstraction to split one big array across multiple nodes. Every node will contain an Array with a contiguous memory buffer belonging to it. After which we create a single ChunkedArray which contains the Arrays of all the nodes. From a user application perspective, we can now use any Arrow supported compute function, and index any data in the array as if it is local. Arrow will resolve an index to a pointer location, which ThymesisFlow will then transparently hand off to the relevant node.

The power here is that we can have columnar data bigger than a single node, without having to share data in between nodes. Data is stored only once, and every node can access the data using its own memory instructions making use of CPU caches. 

\section{Results}\label{sec:results}

Introducing Apache Arrow promises to be a suitable candidate for managing the pitfalls of LC-CSM. However, managing the shared memory introduces some overhead during the allocation and management of objects. In particular:

\begin{itemize}
    \item Communication between nodes to exchange metadata
    \item Cache flushing before and after object initialization
    \item Serialization and de-serialization of table descriptors
    \item Communication during remote memory allocations
\end{itemize}

\begin{table}
\caption{Time to initialize table in remote memory (1GiB of data, uint64 elements). For reference a simple gRPC call takes on average 3.3ms.} \label{tab:benchmark}    \vspace{-0.25cm}
    \centering
    \begin{tabular}{ll}
    \hline
        Component                                  & Time avg {[}ms{]} \\ \hline
        Malloc request (gRPC)                      & 4.99          \\
        Remote pre-write flush (gRPC call + flush) & 51.84         \\
        Write to remote memory                     & 180          \\
        Flush local write cache to remote          & 60.32         \\
        Serialize table descriptor                 & 0.058         \\
        Send table descriptor to other nodes       & 3.23          \\ \hline
        Total                                           & 300.44 \\
        \hline
    \end{tabular}
        \vspace{-0.25cm}
\end{table}%

However, we believe these effects are outweighed by the scalability benefits gained by eliminating transfers of large datasets over the network. Specifically we measured the initialization times as seen in Table~\ref{tab:benchmark}. The total time it takes to create a 1GiB table in remote memory takes $300.44ms$ on average, of which $118ms$ is overhead. Flushing the cachelines of every node takes the longest, as the flushing is done on a per 128-byte cacheline basis, and every flushed line potentially needs to be written to remote memory.

\begin{figure}
    \centering
    \includegraphics[width=\linewidth]{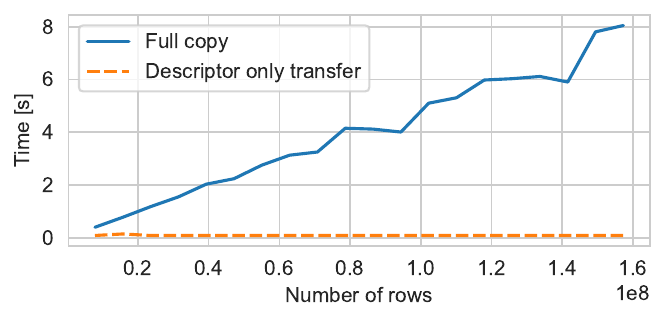}\vspace{-0.5cm}
    \caption{Throughput of data transfers over ethernet compared to sharing metadata zero-copy through the extended Apache Arrow interface.}
    \label{fig:transferthroughput}
\end{figure}

In our LC-CSM testbed built upon a ThymesisFlow installation, we compared our implementation to a standard ethernet transfer of shared data in a cluster. Figure~\ref{fig:transferthroughput} shows that the time spent copying a dataset over ethernet is eliminated by utilizing the disaggregated memory. Only the metadata needs to be transferred over the ethernet link, which allows orders of magnitude faster sharing between nodes.

\begin{figure}
    \centering
    \includegraphics[width=\linewidth]{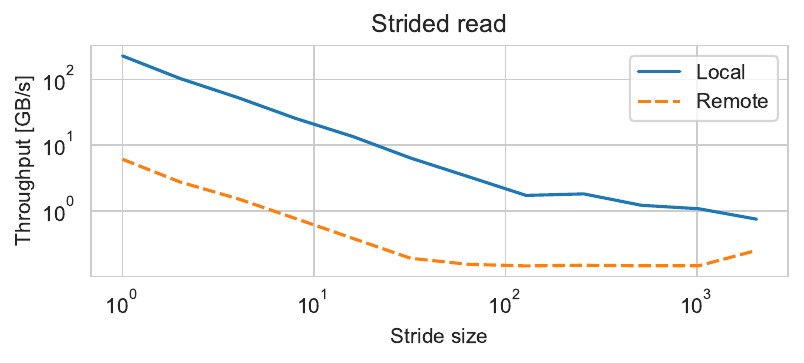}\vspace{-0.5cm}
    \caption{Apache Arrow application throughput for strided read accesses to local and remote memory.}
    \label{fig:strided_read}
\vspace{-0.5cm}
\end{figure}

After the transfer of the metadata, the data is fully and transparently available to the application with a cache-line granularity over the ThymesisFlow link, albeit with a penalty to memory latency and throughput, which becomes the limiting factor. To quantify the impact of LC-CSM on performance, we ran a series of experiments with strided read and write accesses to the remote data, shown in Figure~\ref{fig:strided_read}.
These strided access patterns are common for data analytics pipelines and are a good indicator for the performance penalty incurred by accessing remote memory~\cite{zaidpaper}.
An expected decrease in throughput is measured for an increase in stride size. We theorize this is because more 128-byte transactions need to be done, saturating the ThymesisFlow bus quicker, but also the memory pre-fetching units in the processor may be less able to predict future memory accesses. Comparing the remote to local memory accesses show that remote memory is considerable more limited in maximum throughput, although for a very large stride the difference is considerably less.

In conclusion, we have shown that it is feasible to utilize Apache Arrow for bridging the gap between non-cache-coherent nodes in a LC-CSM setup, maintaining memory consistency as well as flexibility of implementation. While the Arrow interface as well as the ThymesisFlow link do pose some restrictions regarding efficiency, overhead and throughput, we have shown that Arrow is well equipped to become a stepping stone towards facilitating the sharing of large data sets in a shared memory cluster.

\section*{Acknowledgment}

This research was performed with the support of the Eureka Xecs project TASTI (grant no.~2022005).

\bibliographystyle{ACM-Reference-Format}
\bibliography{references}

\end{document}